\documentclass[twocolumn,floatfix]{revtex4-1}
\usepackage{mathrsfs}
\usepackage{physics}
\usepackage{tikz}
\usepackage[caption=false]{subfig}
\usepackage[colorlinks=true,linkcolor=blue,citecolor=blue,urlcolor=blue]{hyperref}
\usepackage{cleveref}
\usepackage{amsmath,amssymb,amsfonts,amsthm}
\usepackage{graphicx}
\usepackage{ragged2e}
\usepackage{float}

\usepackage[normalem]{ulem}

\begin{document}
\date{\today}
\author{Nikolaos K. Kollas}
\email{kollas@upatras.gr}
\affiliation{Division of Theoretical and Mathematical Physics, Astronomy and Astrophysics, Department of Physics, University of Patras, 26504,  Patras, Greece}
\author{Dimitris Moustos}
\email{dmoustos@upatras.gr}
\affiliation{Division of Theoretical and Mathematical Physics, Astronomy and Astrophysics, Department of Physics, University of Patras, 26504,  Patras, Greece}
\author{Kostas Blekos}
\email{mplekos@physics.upatras.gr}
\affiliation{Division of Theoretical and Mathematical Physics, Astronomy and Astrophysics, Department of Physics, University of Patras, 26504,  Patras, Greece}
\title{Field assisted extraction and swelling of quantum coherence for moving Unruh-DeWitt detectors}
\begin{abstract}
We study the effects of motion for an Unruh-DeWitt detector, modeled as a two-level system, on the amount of coherence extracted, when it interacts with a massless scalar coherent field in 1+1 Minkowski spacetime. We observe that compared to a detector at rest, for certain values of the initial energy of the field and the interaction duration, the amount is larger for both a detector moving with a constant speed or uniform acceleration. This ``swelling'' of coherence, which becomes more intense for increasing values of velocity or acceleration, is mostly observed for short durations, when the energy of the field is larger compared to that of the detector, and for longer durations but smaller field energies. As a consequence, the rate at which coherence is lost, is sometimes slower for a moving detector than for a detector at rest.
\end{abstract}
\maketitle
\section{Introduction}
Quantum information science is arguably one of the biggest advances of the 21st century. Although very early on in its development it centered mostly around the properties and applications of entangled systems \cite{RevModPhys.81.865}, recent years has seen a shift in focus towards systems containing \emph{quantum coherence}. This latter property most famously embodied in the notion of a cat both dead and alive inside a box and which according to Feynman ``\textit{lies at the very heart of quantum mechanics}", is a measurable quantity and constitutes a useful resource \cite{PhysRevLett.113.140401,PhysRevLett.116.120404,RevModPhys.89.041003}, which plays a crucial role in a broad range of applications, such as metrology \cite{PhysRevA.94.052324}, biology \cite{Lloyd_2011} and thermodynamics \cite{PhysRevLett.113.150402,Lostaglio2015,PhysRevLett.115.210403,PhysRevX.5.021001,Narasimhachar2015,Korzekwa_2016} to name but a few. On a more fundamental level, it can be shown that coherence and entanglement are actually very closely related, since under certain conditions it is possible to convert one to another \cite{PhysRevLett.115.020403,PhysRevLett.117.020402,PhysRevA.96.032316}.

In this article, we are concerned with the question of extracting coherence onto a quantum system with the help of a quantum field in 1+1 Minkowski spacetime. Motivated by analogous research in the context of  entanglement harvesting, i.e., the process of extracting correlations from the quantum vacuum \cite{VALENTINI1991321,Reznik2003,PhysRevA.71.042104,Salton_2015,PhysRevD.92.064042,PhysRevD.93.044001}, we study the amount of coherence present in an Unruh-DeWitt detector, modeled as a two-level system, after it interacts with a  massless scalar field in a coherent state. Note that even though the general form of the final state of the detector in this case has already been determined in previous research \cite{PhysRevD.96.025020,PhysRevD.96.065008} it nonetheless falls short of any qualitative or quantitative investigations of its coherence and only focuses on the eigenvalues of the detector's density matrix which is independent of the initial choice of coherent field state. 

By choosing a Gaussian coherent amplitude distribution for the field as well as a Gaussian switching function for the interaction, it will be shown that the coherence extracted depends both on the initial energy of the field as well as the interaction duration. By comparing this amount for a detector at rest versus one moving with a constant speed or a uniform acceleration, we observe that for certain choices of the above parameters it is possible in the latter cases to extract a larger amount of coherence, an effect that we call the \emph{swelling of quantum coherence}. As a consequence, for certain values of the initial energy of the field, the rate at which coherence is lost, is sometimes slower if the detector is moving. Thus, this effect could be harnessed as a way of controlling environment induced decoherence.

Due to the fact that the energy of the combined system of the detector plus the field is no longer conserved throughout the process, the existence of coherence in the detector is due not only on the coherence present in the field but also on an incoherent evolution induced by the interaction \cite{PhysRevA.92.032331,BU20171670}.  From a resource theoretic point of view then, the process of coherence extraction resembles less that of entanglement harvesting and more that of assisted distillation of quantum coherence \cite{PhysRevLett.116.070402,PhysRevA.98.052329,Vijayan_2018}.

\section{The model}\label{model}

We consider an Unruh-DeWitt detector \cite{Unruh,DeWitt} interacting with a massless scalar quantum field and moving along a trajectory $x^\mu(\tau)=(t(\tau),x(\tau))$ in 1+1 Minkowski spacetime, with $\tau$ the proper time of the detector.  The detector is modeled as a two-level system (qubit) with frequency $\Omega$. In order to simplify expressions and avoid complications associated with the non-inertial movement of a finite size detector \cite{Schlicht_2004,Louko_2006,PhysRevD.87.064038}, we assume it to be point-like. Throughout, we employ natural units $\hbar=c=1$.

The interaction between the detector and the field is given  by the Unruh-DeWitt Hamiltonian, which in the interaction picture is written as
\begin{equation}
    \hat{H}_{\text{int}}(\tau)=g\chi(\tau)\hat{\mu}(\tau)\hat{\phi}(x^{\mu}(\tau)),
\end{equation}
where $g$ is a small dimensionless coupling constant, and  $\chi(\tau)$ is a non-negative, real valued \emph{switching function}, which specifies how the interaction  between the detector's monopole moment operator $\hat{\mu}(\tau)$ and the field pulled back to the detector's worldline, $\hat{\phi}(x^\mu(\tau))$, is switched on and off. The detector's monopole moment operator is expressed in terms of the ladder operators $\hat{\sigma}_{\pm}$ of the Pauli algebra as
\begin{equation}
    \hat{\mu}(\tau)=e^{i\Omega\tau}\hat{\sigma}_++ e^{-i\Omega\tau}\hat{\sigma}_-,
\end{equation}
while the  field along the detector's trajectory is written as
\begin{equation}
\hat\phi(x^\mu(\tau))=\int\limits_{-\infty}^{+\infty}\frac{dk}{\sqrt{4\pi\abs{k}}}\left[\hat a_ke^{-i\left(\abs{k}t(\tau)-kx(\tau)\right)}+\text{H.c.}\right].
\end{equation}

We now assume that the detector starts out in its ground state and that the initial state of the field is a coherent state $\ket{\alpha}$, defined by
\begin{equation}
    \hat a_k\ket{\alpha}=\alpha(k)\ket{\alpha},
\end{equation}
where $\alpha(k)$ is a coherent amplitude distribution \cite{doi:10.1142/0096}. Employing time dependent perturbation theory, the final state of the detector to second order in the coupling constant is equal to \cite{PhysRevD.96.025020}
\begin{equation}\label{dmatrix}
    \rho=\left(\begin{array}{cc}
         1-\rho_{vac}-\abs{\rho_{coh}}^2& \rho_{coh}\\
        \rho_{coh}^* & \rho_{vac}+\abs{\rho_{coh}}^2
    \end{array}\right)+\mathcal{O}(g^3),
\end{equation}
where $\rho_{vac}$ are vacuum terms, which are independent of the coherent amplitude of the field and
\begin{multline}\label{rcoher}
    \!\!\!\!\rho_{coh}=\!-ig\!\!\!\int\limits_{-\infty}^{+\infty}d\tau\chi(\tau)e^{i\Omega \tau}\!\!\!\int\limits_{-\infty}^{+\infty}\!\!\frac{dk}{\sqrt{4\pi\abs{k}}}\!\left[\alpha(k)e^{-i(\abs{k}t(\tau)-kx(\tau))}\right.\\
    +\left.\alpha^*(k)e^{i(\abs{k}t(\tau)-kx(\tau))}\right].
\end{multline}

A suitable measure, $C$, of the amount of coherence present in Eq. \eqref{dmatrix} is given by the sum of the modulus of its off diagonal elements, known as the \emph{$\ell_1$-norm of coherence} \cite{RevModPhys.89.041003}
\begin{equation}\label{normcoher}
    C=2\abs{\rho_{coh}}+\mathcal{O}(g^3).
\end{equation}

From Eqs. \eqref{rcoher} and \eqref{normcoher} it can be seen that a necessary condition for the presence of coherence in the detector is that the field start out in a coherent state other than the vacuum. Nonetheless, as has already been mentioned, the source of this coherence can no longer traced back to the field alone, but part of it has been created during the interaction. The reason for this is that by construction the Hamiltonian does not conserve the energy of the combined system of detector plus field. A rough analog would be the creation of an equally superposed state when the Hadamard gate acts on either of the computational basis. 

\section{Extraction and swelling of quantum coherence}\label{mainsec}
Suppose we choose the following forms for the switching and amplitude functions
\begin{equation}
    \chi(\tau)=\frac{e^{-\frac{\tau^2}{2T^2}}}{\sqrt{2\pi T^2}},
\end{equation}
\begin{equation}\label{eqa(k)}
    a(k)=\frac{e^{-\frac{k^2}{2E^2}}}{\sqrt{E}},    
\end{equation}
where $T$ is the duration of the interaction and $E$ is the initial energy of the field. 

We now consider the following states of motion for the detector: i) at rest, ii) with constant speed and iii) with uniform acceleration, and study their influence on the amount of coherence extracted. 
\subsection{Detector at rest}
\begin{figure}
    \includegraphics[width=\columnwidth]{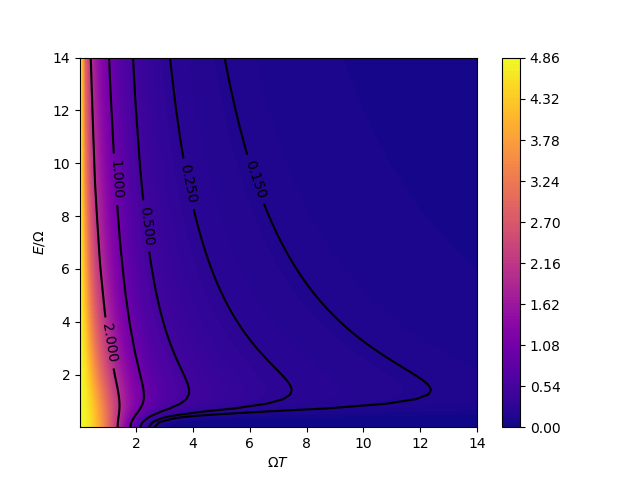}
    \caption{Amount of coherence extracted scaled by the coupling constant, $C_0/g$,  as a function of $E/\Omega$ and $\Omega T$ for a detector at rest. When the energy of the field is of the same order as that of the qubit then coherence decreases more slowly.}
    \label{fig1}
\end{figure}
\begin{figure*}[t]
\begin{minipage}{\textwidth}
\subfloat[$\upsilon=0.5$]{\includegraphics[width=0.25\textwidth]{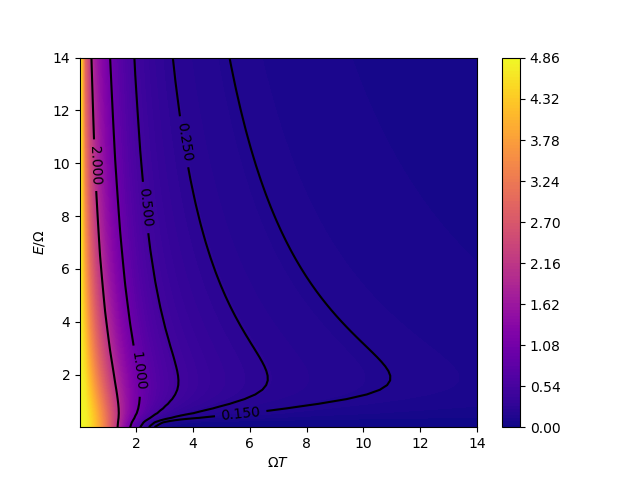}}
\subfloat[$\upsilon=0.7$]{\includegraphics[width=0.25\textwidth]{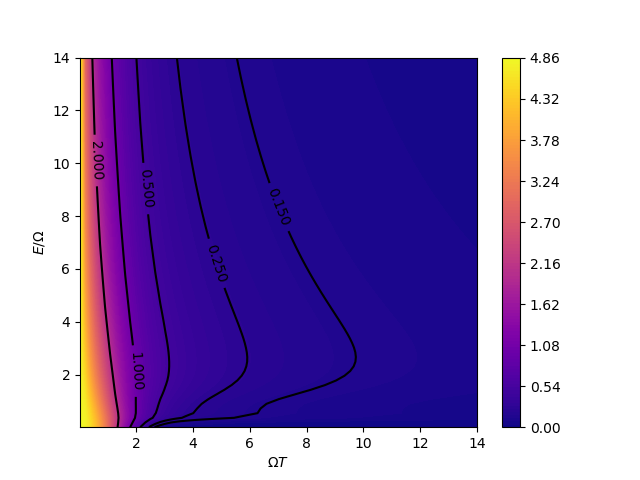}}
\subfloat[$\upsilon=0.8$]{\includegraphics[width=0.25\textwidth]{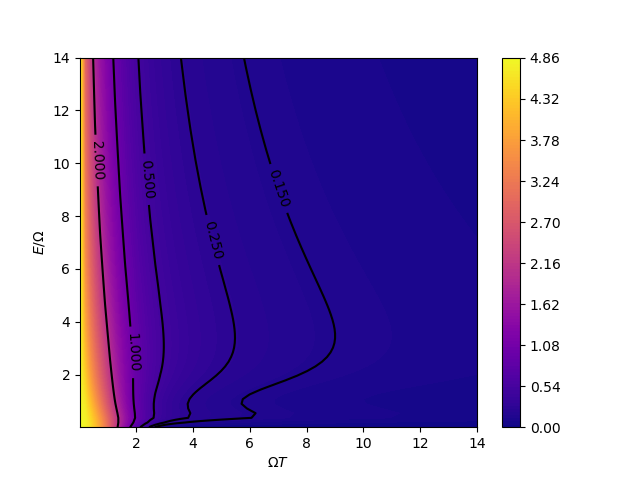}}
\subfloat[$\upsilon=0.99$]{\includegraphics[width=0.25\textwidth]{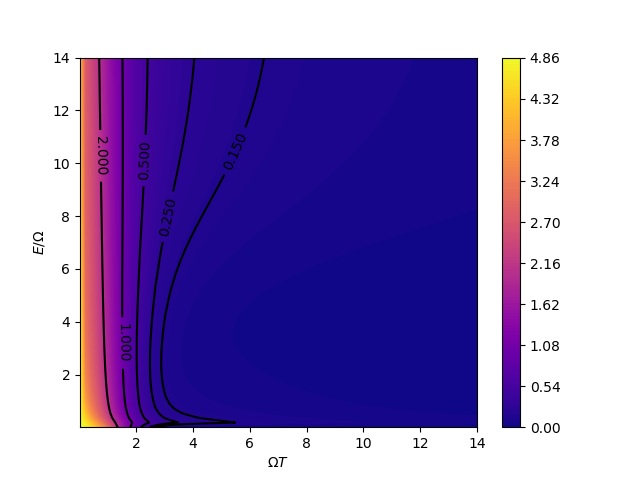}}
\\
\subfloat[$\upsilon=0.5$]{\includegraphics[width=0.25\textwidth]{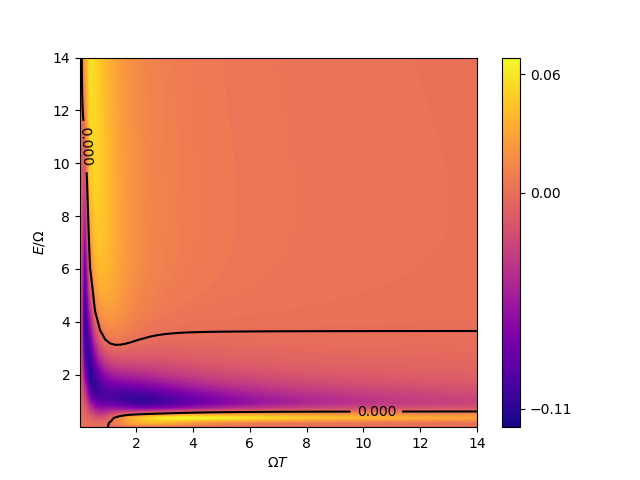}}
\subfloat[$\upsilon=0.7$]{\includegraphics[width=0.25\textwidth]{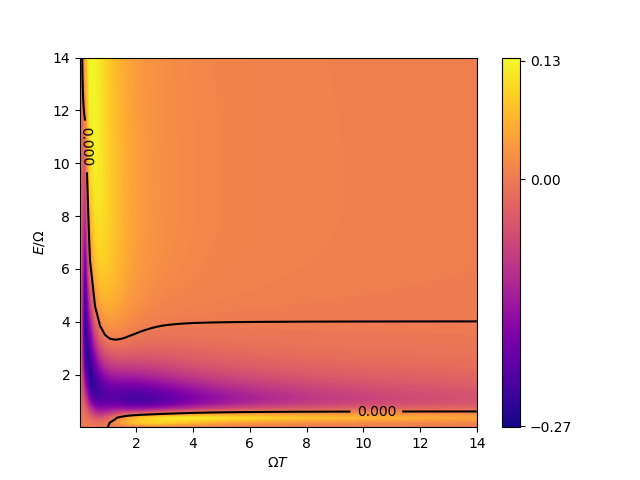}}
\subfloat[$\upsilon=0.8$]{\includegraphics[width=0.25\textwidth]{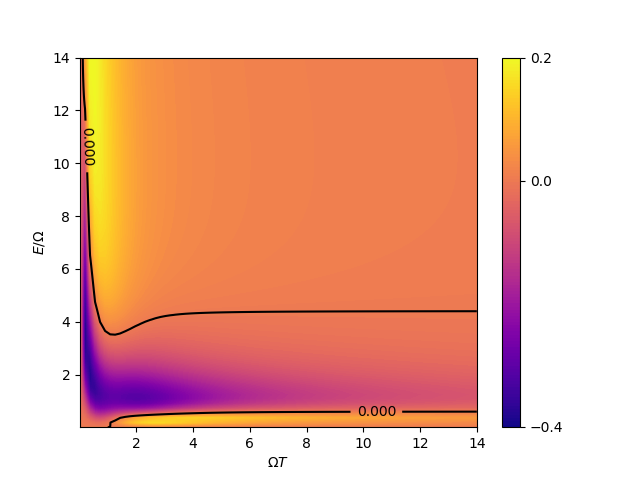}}
\subfloat[$\upsilon=0.99$]{\includegraphics[width=0.25\textwidth]{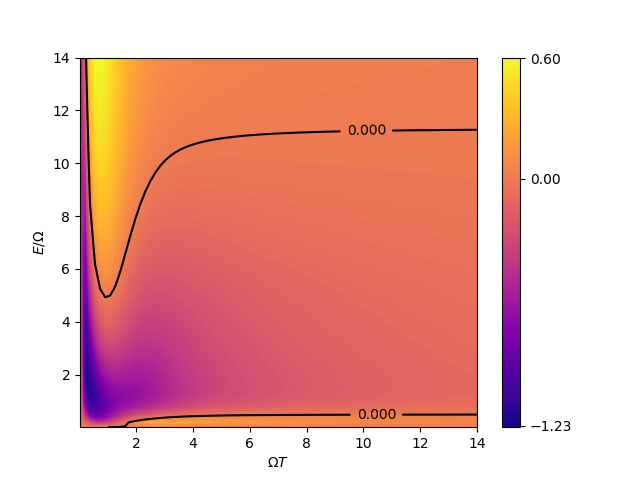}}
\end{minipage}
\caption{(Upper) amount of coherence extracted scaled by the coupling constant, $C_\upsilon/g$,  as a function of $E/\Omega$ and $\Omega T$ for a detector with a constant speed $\upsilon$. (Lower) difference between moving and static case, $C_\upsilon-C_0$. For every value of the detector's speed, there appear ``swelling" regions, in which a moving detector has more coherence than a static one, that are  centered around short interaction duration $T\lesssim 1/\Omega$ when $E>\Omega$ and longer interaction times $T\gtrsim 1/\Omega$ when $E<\Omega$.}
\label{fig2}
\end{figure*}

The spacetime trajectory for a detector at rest at the origin is given by the following coordinates
\begin{equation}
t(\tau)=\tau, \quad x(\tau)=0.
\end{equation}

 Integrating Eq. \eqref{rcoher}, first with respect to $\tau$ and then with respect to $k$, we find that the amount of coherence extracted is given by
\begin{multline}\label{statcoher}
    C_0(\bar{E},\bar{T})=\\g\sqrt{\frac{4\pi \bar{E}\bar{T}^2}{1+\bar{E}^2\bar{T}^2}}e^{-\frac{\bar{T}^2}{2}\left(1-\frac{\bar{E}^2\bar{T}^2}{2(1+\bar{E}^2\bar{T}^2)}\right)}I_{-\frac{1}{4}}\left[\frac{\bar{E}^2\bar{T}^4}{4(1+\bar{E}^2\bar{T}^2)}\right],
\end{multline}
where $I_\nu(z)$ with $\nu\in\mathbb{R}$ is the \emph{modified Bessel function of the first kind} \cite{handbook:math} and we have also introduced the following reduced quantities $\bar{E}=E/\Omega$ and $\bar{T}=\Omega T$. 

In Fig. \ref{fig1}, we present the amount of coherence scaled by the coupling constant, $C_0/g$, for different values of $\bar{E}$ and $\bar{T}$. As one would expect, due to environment induced decoherence, this amount is generally larger for short durations of the interaction, $T\lesssim 1/\Omega$. For a fixed value of the field's initial energy, the amount extracted monotonically decreases, from an initial value of $C_0/g\approx 4.86$. We observe that when the energy of the field is of the same order as that of the qubit, $E\approx\Omega$, then it is possible to maintain coherence for a considerably longer time. 

\subsection{Detector with constant speed}
\begin{figure*}[t]
\begin{minipage}{\textwidth}
\subfloat[$\text{a}/\Omega=0.5$]{\includegraphics[width=0.25\textwidth]{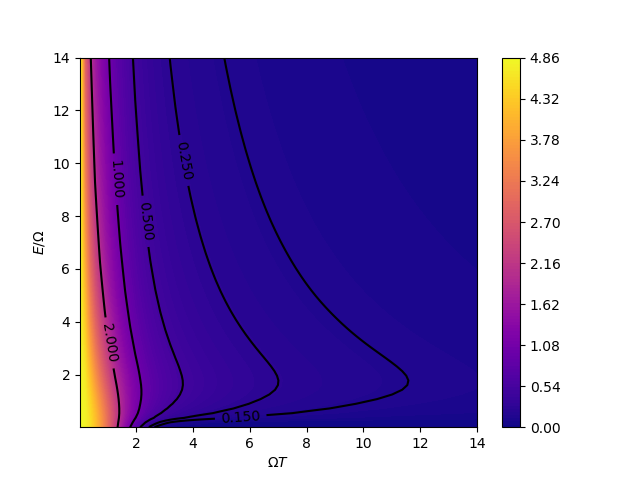}}
\subfloat[$\text{a}/\Omega=1$]{\includegraphics[width=0.25\textwidth]{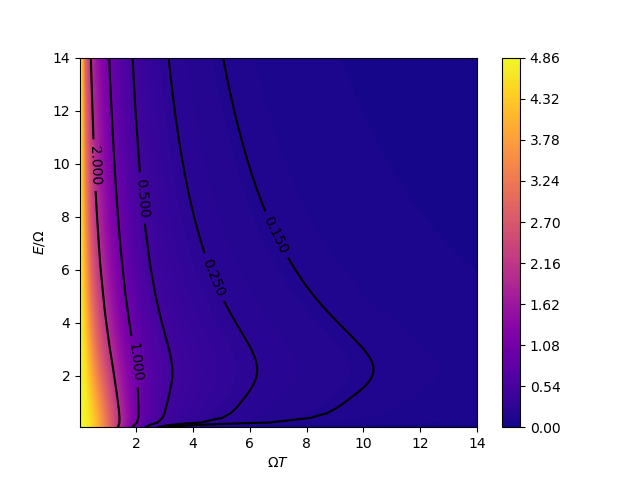}}
\subfloat[$\text{a}/\Omega=2$]{\includegraphics[width=0.25\textwidth]{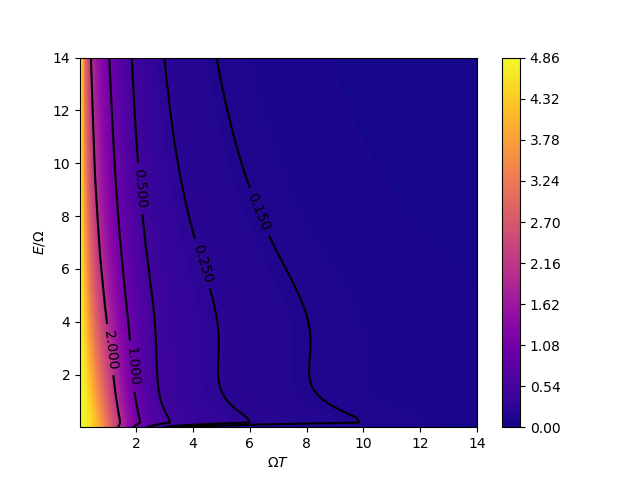}}
\subfloat[$\text{a}/\Omega=5$]{\includegraphics[width=0.25\textwidth]{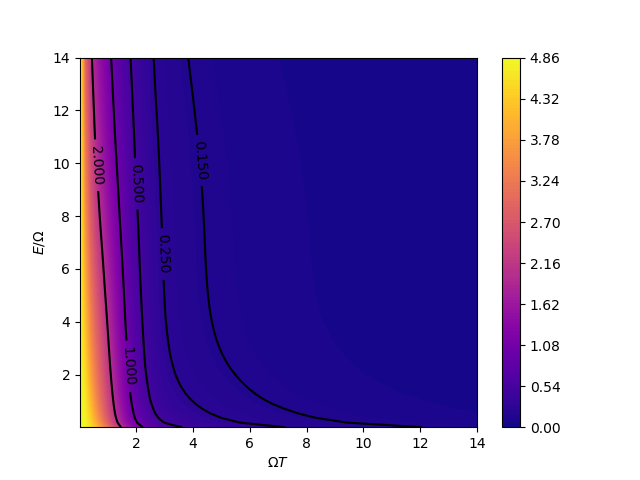}}
\\
\subfloat[$\text{a}/\Omega=0.5$]{\includegraphics[width=0.25\textwidth]{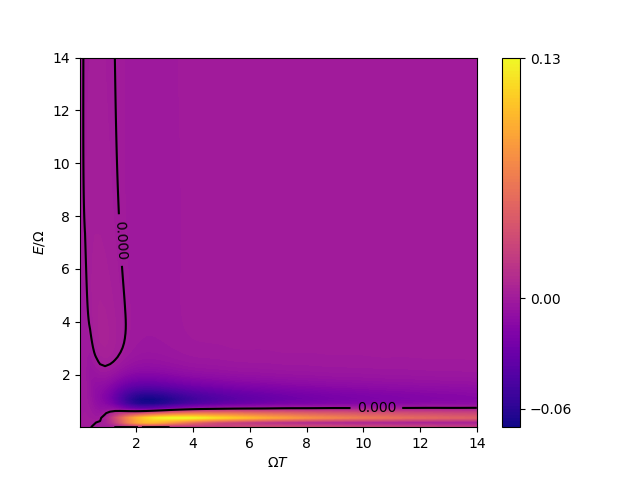}}
\subfloat[$\text{a}/\Omega=1$]{\includegraphics[width=0.25\textwidth]{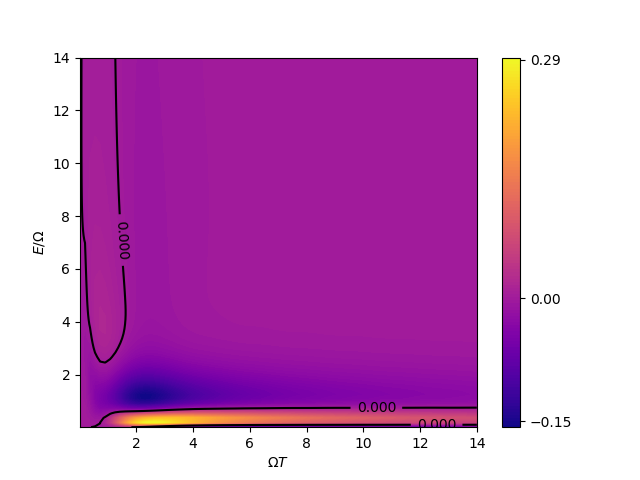}}
\subfloat[$\text{a}/\Omega=2$]{\includegraphics[width=0.25\textwidth]{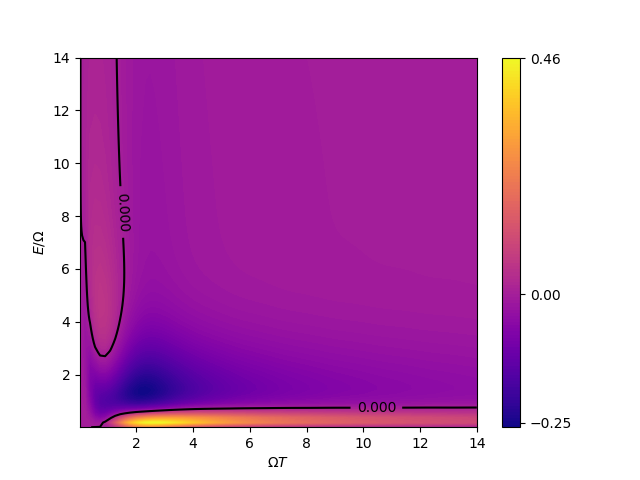}}
\subfloat[$\text{a}/\Omega=5$]{\includegraphics[width=0.25\textwidth]{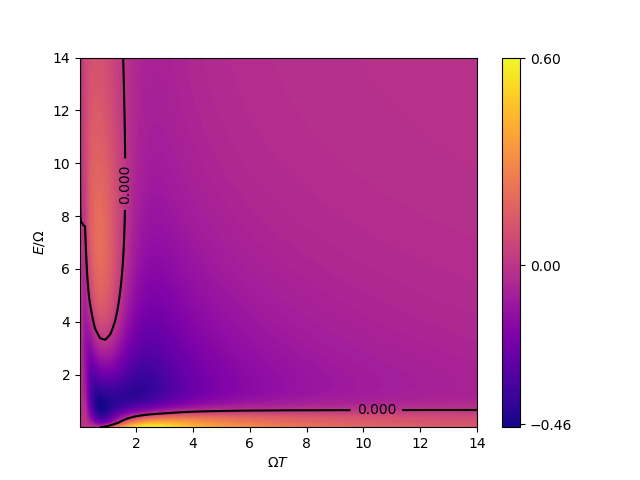}}
\end{minipage}
\caption{(Upper) amount of coherence extracted scaled by the coupling constant, $C_a/g$,  as a function of $E/\Omega$ and $\Omega T$ for a detector with a uniform acceleration $a$. (Lower) difference between moving and static case, $C_\upsilon-C_0$. For every value of the detector's acceleration, there appear only two ``swelling" regions in which a moving detector has more coherence than a static one. One for short interaction duration $T\lesssim 1/\Omega$ when $E>\Omega$ and another one for longer interaction duration $T\gtrsim 1/\Omega$ when $E<\Omega$.}
\label{fig3}
\end{figure*}
Let us now consider a detector following an inertial worldline \cite{birrell}
\begin{equation}
t(\tau)=\gamma \tau, \quad x(\tau)=\gamma \upsilon\tau,
\end{equation}
where $\upsilon$ is the detector's velocity and $\gamma=1/\sqrt{1-\upsilon^2}$ is the Lorentz factor. 

By a simple change of coordinates for $k$ in Eq. (\ref{rcoher}) it can be seen that this case is equivalent to the one for a detector at rest, interacting with a coherent field $\ket{a'}$, with a coherent amplitude equal to an equally weighted convex combination of Eq. (\ref{eqa(k)})
\begin{equation}
    a'(k)=\frac{1}{2}\left(\frac{e^{-\frac{k^2}{2E_{-}^2}}}{\sqrt{E_-}}+\frac{e^{-\frac{k^2}{2E_{+}^2}}}{\sqrt{E_+}}\right),
\end{equation}
with Doppler shifted energies $E_\pm=E\gamma(1\pm\upsilon)$ \cite{SR} and initial energy equal to
\begin{equation}
E_\upsilon=\frac{E}{2}\left(\gamma+\frac{1-\upsilon^2}{1+\upsilon^2}\right).
\end{equation}
In a similar way the coherence extracted is given by the convex combination of the Doppler shifted versions of Eq. (\ref{statcoher}), namely
\begin{equation}\label{eq11}
    C_{\upsilon}(\bar{E},\bar{T})\!=\!\frac{1}{2}\left[C_0\left(\!\bar{E}\sqrt\frac{1-\upsilon}{1+\upsilon},\bar{T}\!\right)\!+C_0\left(\!\bar{E}\sqrt{\frac{1+\upsilon}{1-\upsilon}},\bar{T}\!\right)\right].
\end{equation}

In Fig. \ref{fig2} (a)-(d), we plot Eq. \eqref{eq11} as a function of $\bar{E}$ and $\bar{T}$ for different values of the detector's velocity. We observe that compared to the  detector at rest there are certain regions of the parameter space in which the amount of coherence extracted ``\emph{swells}'' up from it's stationary value. This fact is evident in the positive regions that appear in Fig \ref{fig2} (e)-(h), where we plot the difference, $C_{\upsilon}-C_0$, between the two cases. For increasing values of the detector's speed, this phenomenon becomes increasingly pronounced in the region of short interaction duration,  $T\lesssim 1/\Omega$, when the energy of the field is larger than that of the detector, $E>\Omega$, and the region of long interaction duration, $T\gtrsim 1/\Omega$, for smaller field energies $E<\Omega$.

\subsection{Uniformly accelerated detector}
We consider the case of a linearly, uniformly accelerated detector following the worldline
\begin{equation}
    t(\tau)=\text{a}^{-1}\sinh(\text{a}\tau), \quad x(\tau)=\text{a}^{-1}(\cosh(\text{a}\tau)-1),
\end{equation}
where $\text{a}$ is the detector's proper acceleration \cite{birrell}.
In this case, even though the integration with respect to $k$ can be carried out fairly easily, no closed expression could be found for the integration with respect to the detector's proper time. For this reason, the amount of coherence extracted to a uniformly accelerated detector, $C_a$, cannot be obtained analytically.

In Fig. \ref{fig3}, we numerically evaluate  $C_a$ scaled to the coupling constant as a function of $E/\Omega$ and $\Omega T$ for different values of the detector's acceleration $\text{a}/\Omega$. Similarly to a moving detector with constant speed, swelling of coherence is once again present, and becomes more intense with increasing acceleration. However, in this case the only regions that swelling is observed are those of short and long interaction durations, i.e.,  $T\lesssim 1/\Omega$ for $E>\Omega$ and $T\gtrsim 1/\Omega$ for $E<\Omega$ respectively.
\section{Discussion}\label{concl}
By studying the amount of coherence extracted onto an Unruh-DeWitt detector, which interacts with a massless scalar coherent field, we demonstrated that, to second order in the coupling constant, this amount depends on the initial energy of the field, the duration of the interaction and the state of motion of the detector. As a matter of fact, we have shown that, in the region of short durations, $T\lesssim 1/\Omega$, and large initial field energies, $E>\Omega$, as well as the complementary region of longer durations, $T\gtrsim 1/\Omega$ and small energies, $E<\Omega$, the amount extracted is larger for increasing values of the detector's speed and acceleration.

\begin{figure*}[t]
\begin{minipage}{\textwidth}
\subfloat[$E=0.25\Omega$]{\includegraphics[width=0.5\textwidth]{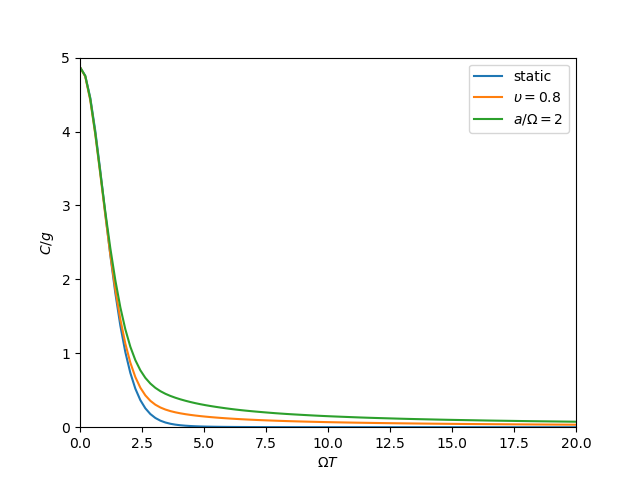}}
\subfloat[$E=\Omega$]{\includegraphics[width=0.5\textwidth]{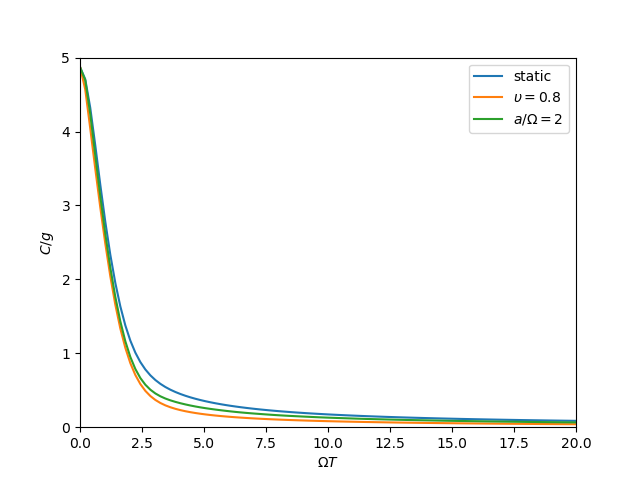}}
\end{minipage}
\caption{Comparison of the amount of extracted coherence scaled by the coupling constant, $g$, between a detector at rest, a moving detector with velocity $\upsilon=0.8$, and a uniformly  accelerated one with acceleration $a/\Omega=2$. Depending on the initial energy of the field, the decoherence rate is sometimes smaller for a moving detector than a detector at rest.}
\label{fig4}
\end{figure*}

The existence or not of swelling, will drastically affect the rate at which coherence is lost. Even though in all three cases, for a fixed value of the initial energy of the field, coherence monotonically decreases with respect to  the duration of the interaction between detector and field, it is the case that when swelling is present, coherence is lost less rapidly for a moving detector than when it is at rest (see Fig. \ref{fig4}). This effect could be exploited as a natural mechanism for the preservation of the detector's coherence against the environment induced decoherence.

In preliminary studies, we have observed that the amount of coherence extracted to a  detector at rest decreases in 3+1 Minkowski spacetime. In  future work \cite{future}, we will present how  swelling is affected in this, and also more realistic scenarios, such as a detector with a finite size and different types of interaction between detector and field.

\acknowledgments{
The authors would like to thank C. Anastopoulos and O. T. Kringer for helpful discussions during preparation of this manuscript.
D. M.'s research is co-financed by Greece and the  European Union (European Social Fund-ESF) through the Operational Programme ``Human Resources Development, Education and Lifelong Learning" in the context of the project ``Reinforcement of Postdoctoral Researchers - 2nd Cycle" (MIS-5033021), implemented by the State Scholarships Foundation (IKY).}
\bibliography{swelling}

\end{document}